# Broken Dynamic Symmetry and Phase Transition Precursor

Yongmei M. Jin,[1]* Yu U. Wang,[1]* Yang Ren,[2] Fengde D. Ma,[1] Jie E. Zhou,[1] Tian-Le Cheng,[1] Ben L. Wang,[1] Liwei D. Geng,[1] Igor A. Kuyanov[1]

**Abstract:** Symmetry breaking is a central concept of Landau phase transition theory, which, however, only considers time-averaged static symmetry of crystal lattice while neglects dynamic symmetry of lattice vibrations thus fails to explain the ubiquitous transformation precursor phenomena. We show that incomplete phonon softening prior to phase transformation leads to dynamic symmetry breaking, whose natural consequences manifest as various precursor "anomalies" that have been difficult to understand from traditional theory. Our experimental observation of heterogeneous phonon domains in high-static-symmetry austenite phase of thermoelastic Ni-Mn-Ga single crystals before martensitic transformation by using three-dimensional synchrotron X-ray phonon diffuse scattering confirms the idea of dynamic symmetry breaking. It provides a natural mechanism and physical understanding of ubiquitous phase transition precursor phenomena in metals, alloys and ceramics.

Martensitic transformation is a typical first-order displacive phase transition in crystalline solids, with fundamental and practical importance to science and technology (*1-3*). Yet, our understanding of this phase transition is still incomplete. For the past half-century it has been known that anomalous precursor phenomena generally occur in metals, alloys and ceramics prior to martensitic transformations (*4-8*). Various anomalies appear in cubic phase in a wide temperature range 10-100 K above martensitic transformation temperature, including diffuse scattering (streaks and satellites) in X-ray, neutron and electron diffraction and cross-hatched nanoscale striation image contrast (tweed patterns) in transmission electron microscopy, which are accompanied by anomalous thermal, acoustic and elastic properties (anisotropic thermal expansion, increased acoustic attenuation, elastic softening, etc) (*9-15*). Because Bragg reflection spots remain those of high-symmetry cubic phase, these phenomena cannot be understood from conventional Landau theory of phase transitions, which only considers time-averaged static symmetry of crystal lattice while neglects the dynamic symmetry of lattice vibrations. Although it has been recognized that phonon softening is essential for understanding martensitic precursor (*4,7*), the physical mechanism and origin of phase transition precursor have remained unclear (*8*).

[1]Department of Materials Science and Engineering, Michigan Technological University, Houghton, MI 49931, USA. [2]X-Ray Science Division, Advanced Photon Source, Argonne National Laboratory, Argonne, IL 60439, USA.
*To whom correspondence should be addressed. E-mail: ymjin@mtu.edu, wangyu@mtu.edu

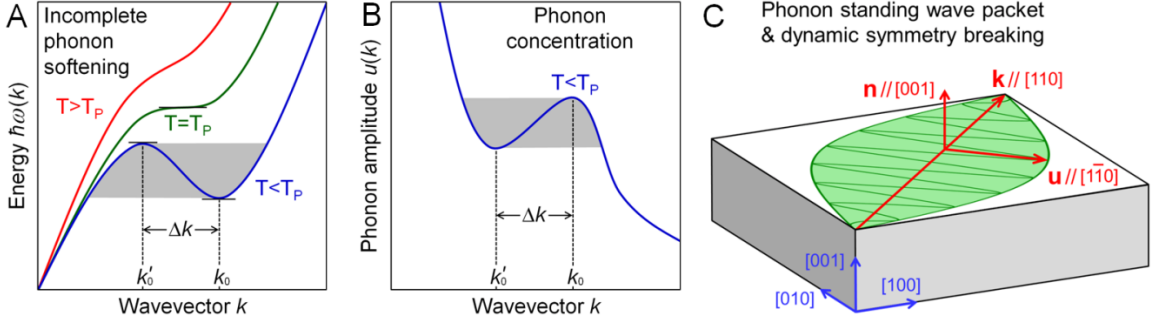

**Fig. 1.** (**A**) Phonon dispersion at temperature $T$ above, equal to, and below $T_P$ (all above $T_M$). Below $T_P$, negative dip develops at wavevector $k_0$, leading to phonon standing wave packets of carrier wavevector $k \approx (k_0 + k_0')/2$ and envelope period $l \approx 2\pi/\Delta k$. (**B**) Number of phonons (phonon amplitude) obeys Bose-Einstein statistics, which at $T<T_P$ concentrates into a peak at wavevector $k_0$ of low-energy phonons. (**C**) Standing wave packet (here [ζζ0] transverse acoustic $TA_2$ mode as example) breaks the dynamic symmetry of lattice vibration among directions of wavevector **k**, polarization **u** and perpendicular **n**.

Here we show that, due to dynamic symmetry breaking prior to the static symmetry breaking of phase transformation, the ubiquitous "anomalous" phase transition precursor phenomena are natural dynamic consequences of incomplete phonon softening, which are the intrinsic properties of anharmonic lattice vibrations and, contrary to previous thought, do not rely on extrinsic defects (*5-8,12,16-18*). To quantitatively corroborate our ideas, we carried out in-situ three-dimensional diffuse scattering experiments on thermoelastic Ni-Mn-Ga single crystals above martensitic transformation temperature by using high-energy synchrotron X-ray diffraction as well as first-principles density functional perturbation theory calculations (*19*). Our ideas are illustrated in Fig. 1. Figure 1A shows phonon dispersion relation $\omega(k)$ at three representative temperatures (all above martensitic transformation temperature $T_M$) with respect to critical temperature $T_P$ during incomplete phonon softening; below $T_P$, negative dip develops at wavevector $k_0$ in the phonon dispersion curve. The thermal average number of phonons $n(k)$ obeys Bose-Einstein statistics, which concentrates at $k_0$ below $T_P$. The phonon amplitude $u(k)$ is related to $n(k)$ through $u^2 \propto (n+½)/\omega$, which accordingly develops a peak at $k_0$ below $T_P$ (Fig. 1B). The intensity distribution $I(k)$ of phonon diffuse scattering from this softened branch depends on phonon amplitude $u$, polarization direction **e**, phonon wavevector **k** and reciprocal lattice vector **G** (corresponding to Bragg reflection) through $I \propto [(\mathbf{G}+\mathbf{k})\cdot\mathbf{e}]^2 u^2(\mathbf{k})$ (*20*), which accordingly concentrates into satellite spot at $k_0$ below $T_P$. The diffuse scattering satellites characteristic of martensitic precursor are produced by low-energy phonons [X-ray diffuse scattering can be used to determine phonon dispersions complementing traditional neutron scattering technique (*21*)]. Such diffuse satellites have been previously treated as superlattice reflection spots associated with static lattice modulations (static atomic displacements) (*5-10,12,15,16*). The apparent incommensurability in satellite position and in traditionally thought static superlattice modulation is indeed associated with phonon wavevector, thus incommensurability is no longer a puzzling issue. Because diffuse scattering of softened transverse phonons (**k**·**e**=0) depends on **G**·**e**, it exhibits systematic extinction. Our experimental evidences are presented in Figs. 2 and 3.

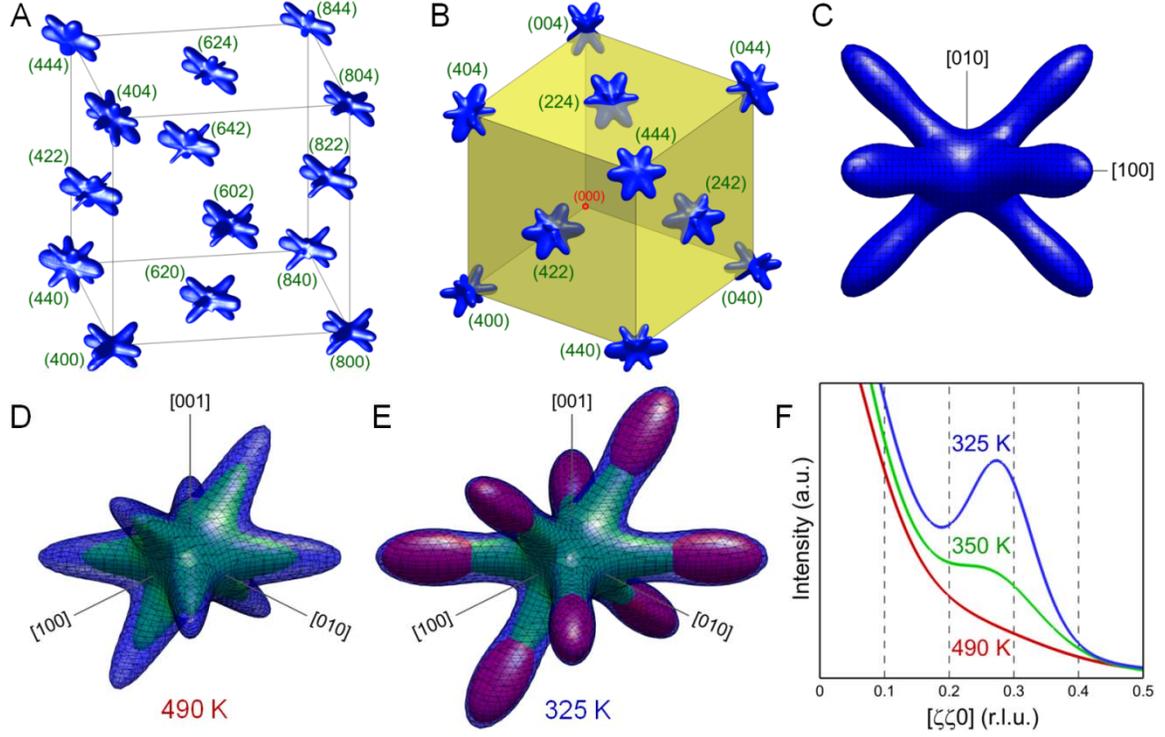

**Fig. 2.** (**A**, **B**) Three-dimensional diffuse scattering of martensitic precursor under stress-free condition (visualized by intensity isosurfaces) at 330K (below $T_P$). Each diffuse rod corresponds to one branch of softened transverse acoustic $TA_2$ phonons with $\langle 110 \rangle$ wavevectors and $\langle 1\bar{1}0 \rangle$ polarizations. Systematic extinction of diffuse rods arises from factor **G**·**e** in diffuse scattering intensity. (**C**) Anisotropic broadening of diffuse scattering rods along polarization **u** and perpendicular **n** directions (Fig. 1C) around (800) Bragg peak. (**D**, **E**) Phonon diffuse scattering at 490K (above $T_P$, without satellites) and 325K (below $T_P$, with satellites) around (800) Bragg peak, as visualized by intensity isosurfaces of the diffuse rods and their constituting component peaks decomposed by peak fitting. (**F**) Temperature dependence of diffuse scattering intensity distribution (arbitrary unit) along one diffuse rod. Diffuse scattering concentrates into satellites upon cooling across $T_P \approx 350K$. The incommensurate position of diffuse satellites gradually shifts towards $\zeta=\frac{1}{3}$ with decreasing temperature but never reach $\zeta=\frac{1}{3}$ at $T_M=323K$.

Concurrent with the development of negative dip in phonon dispersion below $T_P$, phonon standing wave packets of carrier wavevector $k \approx (k_0 + k_0')/2$ and envelope period $l \approx 2\pi/\Delta k$ (Fig. 1, A and B) form in crystal lattice. Electron diffraction probes instantaneous atomic positions of vibrating lattice, producing instantaneous diffraction image contrast. While traveling waves carry such instantaneous image contrast to different regions when moving through the lattice thus averaging out the contrast, standing wave packets consistently produce such diffraction image contrast in same lattice regions that adds up to produce tweed patterns. The unique dip-hump duo in phonon dispersion and phonon population (Fig. 1, A and B, highlighted in grey) lead to pseudo-periodic envelope of phonon standing wave packets. For typical $\Delta k \sim 0.1$ reciprocal lattice unit ($2\pi/a$, $a$ being lattice parameter), the average period is $l \sim 10a$, giving a characteristic length scale of tweeds along wavevector **k** direction. Another length scale of tweeds is associated with the transverse extension $L$ of wavefronts perpendicular to **k**, which produces broadening of diffuse scattering rods (Fig. 2C); the observed anisotropic broadening along polarization **u** and perpendicular **n** directions (Fig. 1C) indicates slightly longer extension along **u** than along **n**, giving $L_\mathbf{u} \sim 25a$ and $L_\mathbf{n} \sim 20a$. These two length scales (width $l$ and length $L$) together with the pseudo-periodicity of standing wave packets lead to characteristic cross-hatched striation tweed patterns, with each set of nanoscale striations perpendicular to the carrier wavevector **k** direction (parallel to the polarization **u** direction in the case of transverse waves). The negative dip in phonon dispersion deepens upon cooling (Fig. 1A), leading to more phonon concentration (Fig. 1B), stronger diffuse satellites (Fig. 2, E and F) and increased diffraction image contrast of tweeds with decreasing temperature. Just like diffuse satellites being misinterpreted as superlattice reflection spots, such tweed patterns have previously been interpreted as indication of static lattice strain stabilized by defects (*5-8,12,16-18*). Such a tradition to seek static structural origin of tweed patterns and transformation precursor follows an early systematic study of tweed diffraction contrast from static shear strains due to coherent precipitation (G.P. zones formed by atomic diffusion) in Cu-Be alloy (*22*). It is worth noting that "tweed" is a description of image contrast rather than a definition of lattice structure, thus the tweeds of martensitic precursor and coherent precipitation must be distinguished.

Unlike traveling waves whose effects are averaged out over time, a phonon standing wave packet produces persistent dynamic effects in the lattice region that it occupies. While the time-averaged atomic positions and thus the static symmetry of the cubic lattice are retained, the dynamic symmetry of the local lattice is broken by the presence of oscillating carrier waves of a single mode under the standing wave packet envelope [transmission electron microscopy confirms that each tweed nanodomain is subjected to a single wavevector (*15*)]. As illustrated in Fig. 1C, for [ζζ0] transverse acoustic TA$_2$ phonon relevant to most martensitic systems, the directions of wavevector **k** along [110], polarization **u** along [1$\bar{1}$0] and perpendicular **n** along [001] are inequivalent; in particular, the **n** direction along [001] is distinct from [100] and [010], which defines a dynamic tetragonal **n**-axis, leading to 6 orientation variants corresponding to 6 crystallographically equivalent TA$_2$ phonon branches (**k** along ⟨110⟩). Therefore, we predict that the lattice regions occupied by phonon standing wave packets behave as anisotropic domains. Since Bragg reflection spots remain those of the cubic lattice manifesting high static symmetry, such heterogeneous phonon domains of broken dynamic symmetry manifest themselves through their responses to external loadings, as confirmed by our synchrotron X-ray diffraction experiments under in-situ stress and first-principles density functional perturbation theory calculations (Fig. 3).

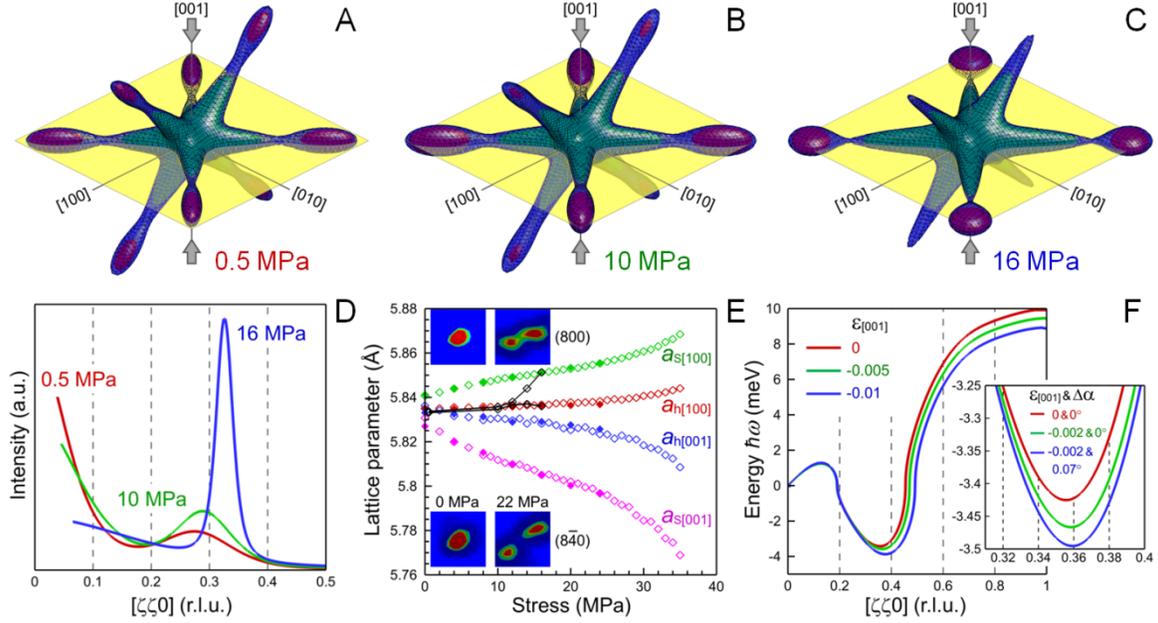

**Fig. 3.** (**A-C**) Phonon diffuse scattering around (800) Bragg peak under [001] stress at 350K (≈$T_P$). (**D**) Stress dependence of diffuse scattering intensity distribution along in-plane diffuse rod (within yellow plane). (**E**) Stress-dependent lattice parameters of hard (red, blue) and soft (green, pink) phonon domains determined from Bragg peaks (insets). Lattice parameters (black) are also determined from centers of diffuse scattering. (**F**) Phonon dispersion in response to [001] stress and shear $\Delta\alpha$ between [100] and [010] axes (Fig. 1C) computed by first-principles density functional perturbation theory.

To detect the heterogeneous phonon domains of broken dynamic symmetry, we performed three-dimensional diffuse scattering experiments under in-situ uniaxial stress applied along single crystal [001] axis at 350K (≈$T_P$). Figure 3A-C shows that the in-plane diffuse scattering rods corresponding to wavevectors **k** perpendicular to [001] stress (**k** along [110] and [1$\bar{1}$0] within yellow plane) grow in intensity with increasing stress, while the out-of-plane diffuse rods are depressed by stress. The in-plane diffuse satellites not only grow in intensity, but also concentrate into narrower peaks (Fig. 3, A-D), and their incommensurate positions gradually shift towards $\zeta$=⅓ with increasing stress (Fig. 3D). These observations indicate similar stress-enhanced phonon softening behaviors, in agreement with neutron scattering experiments (*23,24*). Our first-principles density functional perturbation theory calculations of stress-dependent TA$_2$ branch phonon dispersion (Fig. 3F) further support our experimental observations, in agreement with previous first-principles study (*25*).

More importantly, our experiments show that, while the Bragg peaks remain those of cubic phase under stress-free condition (Fig. 3E, insets) down to $T_M$=323K, they split into two sets under stress (Fig. 3E, insets), corresponding respectively to a soft and a hard tetragonal lattices. The lattice parameters as functions of stress exhibit hysteresis-free reversibility (Fig. 3E, solid-filled symbols for unloading), indicating elastic behaviors of both lattices. The lattice parameters $a_{s[100]}$ and $a_{s[001]}$ respectively along single crystal [100] and [001] axes correspond to the elastically deformed soft phonon domains with dynamic tetragonal **n**-axis (Fig. 1C) parallel to [001] stress (wavevectors **k** along [110], [1$\bar{1}$0]), while $a_{h[100]}$ and $a_{h[001]}$ correspond to the hard domains with **n**-axis along [100] or [010] perpendicular to [001] stress (**k** along [011], [01$\bar{1}$], [101], [10$\bar{1}$]). The split Bragg peak pairs of soft and hard lattices (Fig. 3E, insets) exhibit ~1:2 ratio of integrated intensities, manifesting the volume fractions of the soft and hard heterogeneous phonon domains as determined by the distributions of **k** along ⟨110⟩. While each of these two tetragonal lattices produces own diffuse scattering, the in-plane diffuse scattering is increasingly dominated by the soft phonon domains with increasing stress, and the out-of-plane diffuse scattering is dominated by the hard domains. As a result, the lattice parameter $a$ determined from the center of out-of-plane diffuse scattering around (800) Bragg peak is the same as $a_{h[100]}$, while the lattice parameter $a$ determined from the center of in-plane diffuse scattering is close to $a_{h[100]}$ at low stress and approaches $a_{s[100]}$ with increasing stress (Fig. 3E, black symbols and lines). The hard domains exhibit Young's modulus $E$=14 GPa and Poisson's ratio $\nu$=0.33, while the soft domains exhibit $E$=4.8 GPa and $\nu$=0.47. These phonon domains are significantly softer when compressed along the dynamic tetragonal **n**-axis, and deform elastically with little volume strain ($\nu$≈½); such behavior is mediated by stress-induced redistribution of phonon populations as evidenced by diffuse scattering (Fig. 3, A-D); when compressed in a transverse direction, such phonon domains are significantly harder and exhibit normal Poisson's ratio. These heterogeneous phonon domains are responsible for the anomalies of elastic wave velocity and attenuation observed in premartensitic austenite (*11*). The phonon domains are found to be very sensitive to small stress, and the unavoidable force to mount single crystals onto rotation stage for three-dimensional diffraction tends to induce a small strain (~0.001) to the domains; except for this initial uncertainty, the entire stress-induced strain up to 0.01 is reversible without hysteresis (Fig. 3E). The effects of small stress on the phonon domains explain the anisotropic thermal expansion of premartensitic austenite observed under small applied force (*13,14*) and the "ghost" lattice effect (*9,10*). Furthermore, the broken dynamic symmetry between wavevector **k** and polarization **u** directions (Fig. 1C) also predicts an easy monoclinic shear as characterized by an angle change $\Delta\alpha$ between [100] and [010] axes, which is confirmed by our first-principles density functional perturbation theory calculations (Fig. 3F, inset) but is beyond our current experimental detection capability.

Our results show that dynamic symmetry breaking naturally precedes static symmetry breaking during martensitic transformations. It provides a natural mechanism and physical understanding of the ubiquitous while puzzling phase transition precursor phenomena. Molecular dynamics simulations have shown that dynamic tweeds emerge out of intrinsic anharmonicity without resort to defects (*26,27*). Dynamic symmetry breaking and heterogeneous phonon domains are expected to affect material properties not only in shape memory alloys but also in relaxor ferroelectrics and superconductors that all undergo martensitic-like phase transitions and exhibit precursor phenomena. Phonon domains of broken dynamic symmetry together with the conventional ferroic domains of broken static symmetry enable a complete understanding of displacive (diffusionless) phase transitions and structure-property relations in crystalline solids.

**Acknowledgements:** This work was supported by US DOE Office of Basic Energy Sciences, Materials Sciences and Engineering Division, Physical Behavior of Materials Program under Award No. DE-FG02-09ER46674. Use of the Advanced Photon Source, a US DOE Office of Science User Facility operated by Argonne National Laboratory, was supported by DOE under Contract No. DE-AC02-06CH11357. Density functional theory calculations were performed by using VASP on XSEDE supercomputers.

**Supplementary Materials: Materials and Methods**

**Three-Dimensional Diffuse Scattering Experiments**

Single crystals of Ni$_{49.90}$Mn$_{28.75}$Ga$_{21.35}$ (2×2×4 mm$^3$, {100} orientation, prepared by Adaptamat®) are measured by high-energy synchrotron X-ray diffraction at beamline 11-ID-C of Advanced Photon Source in Argonne National Laboratory. Bulk single crystals are used to facilitate in-situ application of stress. The high energy and high flux of synchrotron X-ray beam are essential for the experiments: high energy (115 keV) provides high penetration depth to probe bulk single crystals, short wavelength (0.107805 Å) enables access to high reciprocal space, and high flux allows short exposure time of each frame necessary for three-dimensional reciprocal space mapping of weak diffuse scattering under in-situ conditions. The experimental setup is calibrated by standard CeO$_2$ powder specimen. A 200×200 μm$^2$ beamsize and 0.1° rotation increment are used to achieve high reciprocal space resolution. Scattering intensity is measured by two-dimensional digital X-ray detector (PerkinElmer® XRD 1621 AN, 2048×2048 pixels, 200 μm pixel size, 16 bit digitization). To prevent signal oversaturation from strong Bragg reflections, Bragg peak intensity is weakened using lead disks positioned in front of the image plate. Exposure time of 6 seconds per frame is used to achieve high signal-to-noise ratio of weak diffuse scattering intensity data. Temperature is controlled in-situ by using cryostream (Oxford Cryosystems® 700 Plus, 80-500 K) without contacting the sample (stress-free condition). A custom-built loading device is used to perform in-situ diffraction under controlled uniaxial stress, which is equipped with loadcell to monitor applied force. Single crystals are glued to the lower loading rod and compressed by the upper loading rod.

Three-dimensional diffuse scattering technique is employed to quantitatively investigate martensitic precursor behaviors. Phonon branches produce multiple diffuse scattering rods, whose intensity distributions are projected onto image plate and overlap in conventional rocking crystal method. Three-dimensional diffraction technique does not suffer such overlapping problem because these diffraction intensities do not overlap in three-dimensional reciprocal space, thus individual diffuse scatterings from different phonon branches are measured.

At each single crystal rotation position during three-dimensional reciprocal space scanning, the image plate measures the two-dimensional diffuse scattering intensity distribution that intersects with the Ewald sphere and is projected onto the flat image plate. The three-dimensional reciprocal space map of diffuse scattering intensity is subsequently constructed from a series of such two-dimensional data recorded by the flat image plate, each of which is first projected back onto the Ewald sphere with the intensity values at individual pixels corrected according to the solid angle subtended by each pixel. The measurement noises can be effectively reduced by pixel averaging (2×2 or 4×4), which also reduces the total amount of three-dimensional digital data and facilitates further quantitative data analysis.

To quantitatively analyze diffuse scattering intensity distributions and Bragg reflection peaks, data fitting code is developed to extract quantitative information. Pseudo-Voigt functions are used to describe the diffuse scattering rods and Bragg peaks. Each diffuse rod is modeled by a combination of 3 three-dimensional pseudo-Voigt peaks, one major peak and two minor peaks located at each end of the diffuse rod. A total of 12 constituting component peaks are used to model the diffuse scattering rods observed around (800) Bragg peak. A least square algorithm is used to converge the data fitting.

**Phonon Dispersion Calculations**

First-principles density functional theory calculations of phonon dispersions are carried out to further support our theoretical ideas and experimental results. Calculations within the framework of density functional theory (DFT) (*28*) and density functional perturbation theory (DFPT) (*29*) are performed to obtain energy-minimized [100] and [010] lattice constants under [001] strain and to determine force constants Hessian matrix, respectively, using Vienna Ab initio Simulation Package (VASP 5.2) (*30*). The plane-wave basis projector augmented-wave (PAW) method (*31*) is used in the framework of spin polarized generalized gradient approximation (GGA) in the Perdew-Burke-Ernzerhof form (*32*). Face-centered cubic $L2_1$ Heusler structure of ordered stoichiometric $Ni_2MnGa$ is considered. The $3d$ electrons of Ga are also included as valence states. For DFT calculations, 500 eV plane-wave cutoff energy and 11×11×8 Monkhorst-Pack $k$-point mesh are used for a tetragonal cell in $(110)_{cubic}$ orientation containing 2 chemical formula units (8 atoms) (*33*). For DFPT calculations, an orthorhombic supercell consisting of 5 tetragonal cells along $[110]_{cubic}$ direction (40 atoms) (*36*) is used with 2×11×8 $k$-point mesh. The phonon frequencies are obtained from the force constants using Phonopy (*34*).